\documentclass[showpacs,amsmath,amssymb,twocolumn,nofootinbib]{revtex4}
\usepackage{epsfig}
\input{epsf}
\usepackage{amssymb}
\begin{document}
\title{Serendipitous discoveries in nonlocal gravity theory}
\author{A.O.Barvinsky}
\affiliation{Theory Department, Lebedev Physics Institute, Leninsky
Prospect 53, 119991 Moscow, Russia}

\begin{abstract}
We present a class of generally covariant nonlocal gravity models which have a flat-space general relativistic (GR) limit and also possess a stable de Sitter (dS) or Anti-de Sitter (AdS) background with an arbitrary value of its cosmological constant. The nonlocal action of the theory is formulated in the Euclidean signature spacetime and is understood as an approximation to the quantum effective action (generating functional of one-particle irreducible diagrams) originating from fundamental quantum gravity theory. Using the known relation between the Schwinger-Keldysh technique for quantum expectation values and the Euclidean quantum field theory we derive from this action the {\em causal} effective equations of motion for mean value of the metric field in the physical Lorentzian-signature spacetime. Thus we show that the (A)dS background of the theory carries as free propagating modes massless gravitons having two polarizations identical to those of the Einstein theory with a cosmological term. The on-shell action  of the theory is vanishing both for the flat-space and (A)dS backgrounds which play the role of stable vacua underlying respectively the ultraviolet and infrared phases of the theory. We also obtain linearized gravitational potentials of compact matter sources and show that in the infrared (A)dS phase their effective gravitational coupling $G_{\rm eff}$ can be essentially different from the Newton gravitational constant $G_N$ of the short-distance GR phase. When $G_{\rm eff}\gg G_N$ the (A)dS phase can be regarded as a strongly coupled infrared modification of Einstein theory not only describing the dark energy mechanism of cosmic acceleration but also simulating the dark matter phenomenon by enhanced gravitational attraction at long distances.
\end{abstract}
\maketitle

\section{Introduction}
Central problem in the attempts to build a modified gravity theory as a model of observable cosmic acceleration \cite{acceleration} consists in elimination of ghost instabilities that usually make any model physically inconsistent. Combined with the necessity to retain the general relativistic limit, this makes the simplest appropriate version of general relativity (GR) -- explicit cosmological term -- very robust against possible attempts to modify it, either by introducing higher-derivative terms or relaxing its diffeomorphism symmetry.

On the other hand, there is fine tuning problem associated with the hierarchy of the fundamental Planck scale vs the cosmic acceleration scale and also the cosmic coincidence problem -- order of magnitude equality of dark energy (DE) and matter contributions (including dark matter (DM)). These problems serve as a strong motivation to go beyond introduction of explicit cosmological term and give rise to numerous models (like $R+R^2/\Lambda$ models \cite{BoulwareDeser}, quintessence \cite{quintessence}, $f(R)$ models \cite{f(R)}, non-minimally coupled matter fields \cite{Shaposhnikovetal}, brane theories \cite{branes}, massive gravity \cite{massive,massive2}, nonlocal cosmology \cite{nonloccosm,DeffWood,Woodard2}), etc.). However, in this or that way fine tuning is creeping into almost all of these models. Modulo certain exceptions \cite{Shaposhnikovetal}, most of them in fact look as a sophisticated way to incorporate into their action in addition to the Planck scale the horizon scale (whether it is explicit cosmological constant, graviton mass of massive gravity \cite{massive}, multi-dimensional Planck mass in braneworld theories or the crossover scale in brane induced gravity models \cite{branes}, etc.).

To circumvent this difficulty one could adopt another, perhaps more promising, line of reasoning. If {\em a concrete fixed scale} incorporated in the model is not satisfactory, then one could look for a model that admits cosmic acceleration scenario with {\em an arbitrary scale}. Its concrete value compatible with observations should arise dynamically by the analogue of symmetry breaking to be considered separately. Even this very unassuming approach is full of difficulties, because modified gravity models featuring this property (like unimodular gravity \cite{unimodular}, $f(R)$-gravity, etc.) generally violate some of its conventional symmetries, have additional degrees of freedom and might lead to ghost instabilities. Thus we get back to the central problem in the modification of Einstein theory, mentioned above, which is especially very actual under the requirement to preserve general covariance of the model.

Here we present a nonlocal infrared modification of Einstein gravity theory briefly reported in \cite{nonloc}, which is likely to implement the above approach. It will be based on the realization of the old idea of a scale-dependent gravitational coupling -- nonlocal Newton constant \cite{AHDDG,covnonloc,HamberWilliams,DvaliHofmannKhoury} -- and will amount to the construction of the class of diffeomorphism invariant models. These models are compatible with the general relativistic (GR) limit and generate a stable de Sitter (dS) or anti-de Sitter (AdS) background with an {\em arbitrary value} of its effective cosmological constant $\Lambda$. Repeating the above motivation again, arbitrariness of $\Lambda$ will be a manifestation of the fact that, to resolve such issues of DE as fine tuning and cosmic coincidence, the scale of $\Lambda$ cannot be encoded in the action of the theory, but rather should arise dynamically by the analogue of symmetry breaking.

In addition to fine-tuning argumentation of the above type, the driving force of our approach will also be the aesthetical motivation to consider as a source of DE a purely metric sector of the theory. No special matter fields like quintessence \cite{quintessence}, dilaton \cite{Shaposhnikovetal} or Lorentz breaking khronon \cite{BlasSibiryakov} will be assumed to exist. In distinction from local $f(R)$ models this will be a {\em nonlocal} generally covariant modification of the metric sector of Einstein theory. Also, it will go beyond the $f(R/\Box)$ models of \cite{nonloccosm,DeffWood,Woodard2} by involving a very nonlinear and nonlocal dependence on all components of Ricci and Riemann tensors. Finally, though our model stems from the idea of a nonlocal gravitational coupling ``constant" \cite{AHDDG} motivated by the idea of degravitation of the vacuum energy \cite{DvaliHofmannKhoury}, no degravitation of matter will be considered, because the source of the effective cosmological constant will be entirely the metric sector of the theory.

Although the main goal of this approach was the achievement of DE phenomenon, the serendipitous nature of the resulting model will appear as an unexpected bonus -- dark matter simulation also generated by the metric sector of the theory. Whatever speculative is the interpretation of this effect, it arises as an enhancement of the gravitational attraction of the ordinary (non-dark) matter at large distances, which is quite opposite to the degravitation mechanism.

The plan of the paper is as follows. In Sect.II we formulate the four-parameter family of nonlocal gravitational models and give a brief summary of their properties -- existence of a stable flat-space background corresponding to their GR limit and an alternative ghost-free (A)dS background with an arbitrary value of the effective $\Lambda$, corresponding to their infrared DE phase. Here we also give the expression for the linearized gravitational potential of matter sources in this (A)dS phase, which has a DM interpretation. In Sect.III within a flat-space perturbation theory setup we give a motivation for the nonlocal structure of their action, based on the generally covariant concept of the scale-dependent gravitational coupling and the requirement of stability with respect to ghost modes. In Sect.IV we discuss treatment of nonlocality under the assumption that the suggested nonlocal action is in fact a certain approximation for the quantum effective action obtained from some fundamental quantum gravity theory. Here we employ the relation between the Schwinger-Keldysh technique for expectation values \cite{SchwKeld} and the Euclidean quantum field theory to show how a nonlocal effective action leads to causal retarded nature of nonlocal equations of motion for mean fields \cite{beyond}. We begin with this relation within the flat-space perturbation theory setup, but then show that in the cosmological context this setup fails unless we modify nonlocal structures of the effective action by extra non-polynomial dependence on the spacetime curvature. In Sect.V we establish this dependence in the action (thus recovering its final nonlocal form introduced in Sect.II) and prove the existence of the (A)dS background with an arbitrary $\Lambda$. Stability requirement for this background requires knowledge of the quadratic part of the action. This represents a technically simple but very lengthy calculational challenge accomplished in this section and paper appendix. In Sect.VI we show that this (A)dS background carries as free propagating modes a massless graviton with two polarizations and also obtain the linearized gravitational potentials of matter sources, which is likely to simulate the effect of DM. Finally, in Sect.VII we discuss GR and (A)dS phases of our theory and the extent to which it might match with the cosmological concordance model and also its prospective nature in other ramifications of quantum gravity theory.

\section{Summary of results}
The action we will be interested in reads in its full complexity as the following nonlocal diffeomorphism invariant functional of the spacetime metric $g_{\mu\nu}$,\footnote{We use the Euclidean signature spacetime and curvature tensor conventions, $R=g^{\mu\nu}R_{\mu\nu}=g^{\mu\nu}R^\alpha_{\;\;\mu\alpha\nu}= g^{\mu\nu}\partial_\alpha\Gamma^\alpha_{\nu\mu}
-...\;$.}
    \begin{eqnarray}
    &&S=\frac{M^2}2\int dx\,g^{1/2}\,\left\{-R+
    \alpha\,R^{\mu\nu}
    \frac1{\Box+\hat P}\,G_{\mu\nu}
    \right\},\;\;\;\;                        \label{action}\\
    &&\hat P\equiv P_{\alpha\beta}^{\;\;\;\mu\nu}
    =a R_{(\alpha\;\;\beta)}^{\;\;\,(\mu\;\;\,\nu)}
    +b \big(g_{\alpha\beta}R^{\mu\nu}
    +g^{\mu\nu}R_{\alpha\beta}\big)               \nonumber\\
    &&\qquad\qquad\qquad
    +c R^{(\mu}_{(\alpha}\delta^{\nu)}_{\beta)}
    +d R\,g_{\alpha\beta}g^{\mu\nu}
    +e R \delta^{\mu\nu}_{\alpha\beta}.     \label{potential}
    \end{eqnarray}
Here $G_{\mu\nu}=R_{\mu\nu}-\frac12g_{\mu\nu}R$ is the Einstein tensor, the hat denotes the matrix acting on symmetric tensors and we use the condensed notation for the Green's function of the operator
    \begin{eqnarray}
    \Box+\hat P\equiv\Box\,\delta_{\alpha\beta}^{\;\;\;\mu\nu}
    +P_{\alpha\beta}^{\;\;\;\mu\nu},
    \quad\Box=g^{\lambda\sigma}\nabla_\lambda\nabla_\sigma,
    \end{eqnarray}
acting on any symmetric tensor field $\Phi_{\mu\nu}$ as
    \begin{eqnarray}
    &&\frac1{\Box+\hat P}\,\Phi_{\mu\nu}(x)\equiv
    \Big[\,\frac1{\Box+\hat P}\,\Big]_{\mu\nu}^{\alpha\beta}\Phi_{\alpha\beta}(x)\nonumber\\
    &&\qquad\qquad\qquad\quad=\int dy\,G_{\mu\nu}^{\alpha\beta}(x,y)\,\Phi_{\alpha\beta}(y)
    \end{eqnarray}
with $G_{\mu\nu}^{\alpha\beta}(x,y)$ -- the two-point kernel of this Green's function.

The boundary conditions for this Green's function will be discussed in much detail in Sects.III-IV. Here we only say that the action (\ref{action}) is formulated in the Euclidean signature spacetime, and in the flat-space background setup it is understood as metric perturbation expansion on this background with trivial zero boundary conditions at infinity which uniquely fix the zeroth order Green's function $1/\Box$. Formal resummation of this expansion series allows one to go over from the flat-space expansion to the expansion on maximally symmetric de Sitter (dS) or Anti-de Sitter (AdS) background with the Green's function uniquely defined either by the condition of regularity on $S^4$ (dS case) or boundary conditions for the Euclidean AdS spacetime corresponding to the definition of the Hartle-Hawking vacuum. As the question of our primary interest here will be the existence of these two vacua (flat-space vacuum and (A)dS one), this information will be sufficient to specify the Green's function, for which we will only require the following symmetric variational law (with respect to local metric variations in $\Box$ and $\hat P$)
    \begin{eqnarray}
    \delta\frac1{\Box+\hat P}=-\frac1{\Box+\hat P}\,\delta\big(\Box+\hat P\big)
    \frac1{\Box+\hat P},              \label{symvar}
    \end{eqnarray}
characteristic of the Euclidean signature d'Alembertian with zero boundary conditions. The relevance of this Euclidean space setup to real physical setting in Lorentzian spacetime and, in particular, to causal effective equations with retarded nonlocalities will be discussed in Sect.IV below.

The action (\ref{action}) has one dimensional parameter $M$ and depends on six dimensionless parameters $\alpha$, $a$, $b$, $c$, $d$ and $e$, the first one $\alpha$ determining the overall magnitude of the nonlocal correction to the Einstein term and the rest -- entering the potential term (\ref{potential}) of the operator $\Box+\hat P$. For a small value of $\alpha$ and the value of $M$ related to the Planck mass $M_P$,
    \begin{eqnarray}
    &&|\alpha|\ll 1,\\
    &&M^2=\frac{M^2_P}{1-\alpha},  \label{M_Prenorm}
    \end{eqnarray}
the theory (\ref{action}) has a GR limit on a flat-space background, whereas the rest of the parameters should be restricted by the requirement of a stable (A)dS solution existing in this theory, which in the dS case can serve as a DE model of accelerating cosmic expansion.

The main claim of our paper is that the theory (\ref{action}) has a stable (ghost-free) de Sitter ($\Lambda>0$) or Anti-de Sitter ($\Lambda<0$) solution of its variational equations of motion with an {\em arbitrary} $\Lambda$, provided its dimensionless parameters satisfy the following restrictions
    \begin{eqnarray}
    &&\alpha=-A-4B,    \label{relation}\\
    &&C=\frac23,   \label{Crelation}
    \end{eqnarray}
where the new set of capitalized quantities $A$, $B$ and $C$ read as
    \begin{eqnarray}
    &&A=a+4\,b+c,                     \label{A}\\
    &&B=b+4\,d+e,                     \label{B}\\
    &&C=\frac{a}3-c-4e.            \label{C}
    \end{eqnarray}
These quantities arise in the coefficients of two tensor projectors on traceful and traceless subspaces, which remain in the potential term (\ref{potential}),
    \begin{eqnarray}
    &&P_{\alpha\beta}^{\;\;\mu\nu}\,\big|_{\,\rm (A)dS}
    =\frac{A+4B}4\,\Lambda\,
    g_{\alpha\beta}g^{\mu\nu}\nonumber\\
    &&\qquad\qquad\qquad
    -C\,\Lambda\Big(\delta^{\mu\nu}_{\alpha\beta}
    -\frac14\,
    g_{\alpha\beta}g^{\mu\nu}\!\Big),      \label{PonAdS}
    \end{eqnarray}
after its calculation on the (A)dS background with the Riemann and Ricci tensors of the form
    \begin{eqnarray}
     &&R_{\alpha\mu\beta\nu}=
     \frac{\Lambda}3\,(g_{\alpha\beta} g_{\mu\nu}-g_{\alpha\nu}g_{\beta\mu}),    \label{Riemann}\\
     &&R_{\mu\nu}=\Lambda\,g_{\mu\nu}.         \label{Ricci}
    \end{eqnarray}

The condition (\ref{relation}) guarantees the existence of the (A)dS solution, while Eq.(\ref{Crelation}) is responsible for its stability. This fact, which can be regarded as a major technical achievement of this work, is based on the calculation of the quadratic part of the action on its (A)dS background. Bearing in mind generality and complexity of the original multi-parameter action, the result turns out to be remarkably simple. When the metric perturbations $h_{\mu\nu}$ on this background are subject to the {\em background covariant} DeWitt gauge condition,
    \begin{eqnarray}
    \chi^\mu\equiv\nabla_\nu
    h^{\mu\nu}-\frac12\,\nabla^\mu h=0,  \label{DWgauge}
    \end{eqnarray}
($h\equiv g^{\mu\nu}h_{\mu\nu}$ and the indices of $h_{\mu\nu}$ are raised by the background metric) this quadratic part depends only on the traceless part of $h_{\mu\nu}$ and reads
    \begin{eqnarray}
    &&S_{(2)}=\frac{M^2_{\rm eff}}2
    \int d^4x\,g^{1/2}\left\{-\frac14\,\bar h^{\mu\nu}\Box\, \bar h_{\mu\nu}\right.
    \nonumber\\
    &&\qquad\qquad
    -\frac14\,\left(C-\frac43\right)\,\Lambda\,\bar h_{\mu\nu}^2
    \nonumber\\
    &&\qquad\qquad
    -\frac{\Lambda^2}4\,
    \left(C-\frac23\right)^2\,\left.
    \bar h^{\mu\nu}\frac1{\Box
    -C\Lambda}\,\bar h_{\mu\nu}
    \right\},                                 \label{quadrS}\\
    &&\bar h_{\mu\nu}
    \equiv h_{\mu\nu}-\frac14\,g_{\mu\nu}h.  \label{traceless}
    \end{eqnarray}
Here the condition (\ref{Crelation}) signifies the absence of the nonlocal term and guarantees that the characteristic equation for all propagating modes will have only one ``massless" root for $\Box$, $\Box=2\Lambda/3$ (on-shell condition of masslessness for (A)dS background differs from its flat-space analogue, $\Box=0$ \cite{AllenTseytlinAdS}). Therefore longitudinal and trace modes which formally have a ghost nature become unphysical and can be eliminated by residual gauge transformations preserving the gauge (\ref{DWgauge}). This well-known mechanism leaves us with two transverse-traceless physical modes propagating on the (A)dS background -- absolutely the same situation as in the Einstein theory with two graviton modes on a flat or (A)dS spacetime.

What is critically different from the GR phase of the theory -- the effective Planck mass $M_{\rm eff}$ in (\ref{quadrS}), which determines the cutoff scale of perturbation theory in the (A)dS phase and the strength of the gravitational interaction of matter sources. It is given by
    \begin{eqnarray}
    &&M_{\rm eff}^2
    =\frac{8B\,(2B+\alpha)}{\alpha\,(1-\alpha)}
    \,M^2_P.                                 \label{effectivemass} \end{eqnarray}
In the presence of matter sources with the stress tensor $T_{\mu\nu}$ the theory on the (A)dS background has linearized gravitational potentials which equal modulo the gauge transformation
    \begin{eqnarray}
    &&h_{\mu\nu}
    =\frac{8\pi G_{\rm eff}}{-\Box+\frac23\Lambda}\,\left(T_{\mu\nu}
    +g_{\mu\nu}\,
    \frac{\Box-2\Lambda}{\Box+2\Lambda}\,
    \frac{\Lambda}{3\Box}\,T\right).     \label{matpot0}
    \end{eqnarray}
Here all nonlocal operations should be understood with retarded boundary conditions (see discussion below in Sects. IV and VI) and
$G_{\rm eff}\equiv 1/8\pi M_{\rm eff}^2$ is the effective gravitational constant vs the Newton constant $G_N=1/8\pi M_P^2$.

For a wide range of free parameters of our model $G_{\rm eff}$ can be much larger than $G_N$ because in view of (\ref{relation}) a natural range of the parameter $B$ is $B\sim\alpha$, and $G_{\rm eff}\sim G_N/\alpha\gg G_N$. This property can be interpreted as a simulation of DM mechanism, because it implies strengthening of the gravitational attraction in the (A)dS phase of the theory and its possible effect on rotation curves at relevant distance scales.

\section{Scale dependent coupling and flat-space background setup}
The choice of the action (\ref{action}) might look contrived, but we will show now that it naturally arises within the concept of the effective scale-dependent gravitational constant. At a qualitative level this concept was introduced in \cite{AHDDG} as an implementation of the idea that the effective cosmological constant in modern cosmology is very small not because the vacuum energy of quantum fields is so small, but rather because it gravitates too little. This degravitation is possible if the effective gravitational coupling constant depends on the momentum scale and becomes small for fields nearly homogeneous at the horizon scale. Naive replacement of the Newton constant by a nonlocal operator suggested in \cite{AHDDG} violates diffeomorphism invariance, but this procedure can be done covariantly due to the following observation \cite{covnonloc}.

The Einstein action in the vicinity of a flat-space background can be rewritten in the form
    \begin{equation}
    S_E=
    \frac{M_P^2}2\int dx\,g^{1/2}\,\left\{\,
    -R^{\mu\nu}\frac1{\Box}\,G_{\mu\nu}\,
    +{\rm O}\,[R_{\mu\nu}^3]\,\right\},       \label{flatE}
    \end{equation}
where $1/\Box$ is the Green's function of the covariant d'Alembertian acting on a symmetric tensor. This expression is nothing but a generally covariant version of the quadratic part of the Einstein action in metric perturbations $h_{\mu\nu}$ on a flat-space background. When rewritten in terms of the Ricchi tensor $R_{\mu\nu}\sim \nabla\nabla h+O[h^2]$ this expression becomes nonlocal but preserves diffeomorphism invariance to all orders of its curvature expansion. In its turn the quadratic nature of the Einstein action in the vicinity of a flat-space background follows from the subtraction of the linear in $h_{\mu\nu}$ term by the surface Gibbons-Hawking integral over asymptotically-flat infinity
    \begin{eqnarray}
    &&S_E=-\frac{M_P^2}2
    \int dx\,g^{1/2}\,R(\,g\,)\nonumber\\
    &&\qquad\qquad+\frac{M_P^2}2
    \int_\infty d\sigma^\mu\,
    \big(\partial^\nu
    h_{\mu\nu}-\partial_\mu h).           \label{GHsubtraction}
    \end{eqnarray}
From the viewpoint of the metric in the interior of spacetime this surface term is a topological invariant depending only on the asymptotic behavior $g^\infty_{\mu\nu}=\delta_{\mu\nu}+
h_{\mu\nu}(x)\,|_{\,|x|\to\infty}$.  It can be converted into the form of the volume integral and covariantly expanded in powers of
the curvature. This expansion starts with\footnote{Covariant way to check this relation is to calculate the metric variation of this integral and show that its integrand is the total divergence which yields the surface term of the above type linear in $\delta
g_{\mu\nu}(x)=h_{\mu\nu}(x)$ \cite{TMF}.}
    \begin{eqnarray}
    &&\int_\infty\! d\sigma^\mu\,
    \big(\partial^\nu
    h_{\mu\nu}-\partial_\mu h\Big)\nonumber\\
    &&\quad\quad=
    \int dx\,g^{1/2}\left\{R
    -R^{\mu\nu}\frac1{\Box}\,G_{\mu\nu}
    +{\rm O}\,[\,R_{\mu\nu}^3\,]\right\},     \label{GH}
    \end{eqnarray}
so that the Ricci scalar term gets canceled in (\ref{GHsubtraction}) and we come to (\ref{flatE}).

With this new representation of the Einstein action, the idea of a nonlocal scale dependent Planck mass \cite{AHDDG} can be realized as the replacement of $M_P^2$ in (\ref{flatE}) by a nonlocal operator -- a function $M^2(\Box)$ of $\Box$,
    \begin{equation}
    M_P^2 R^{\mu\nu}\frac1\Box\,G_{\mu\nu}\Rightarrow
    R^{\mu\nu}\frac{M^2(\Box)}{\Box}\,G_{\mu\nu},
    \end{equation}
which would realize this idea at least within the lowest order of the covariant curvature expansion. This modification put forward in \cite{AHDDG,covnonloc} did not, however, find interesting applications because it has left unanswered a critical question -- is this construction free of ghost instabilities for any nontrivial choice of $M^2(\Box)$? Here we try to fill up this omission and put some constraints on $M^2(\Box)$.

To begin with, if we adopt this strategy, then the search for $M^2(\Box)$ should be encompassed by the correspondence principle according to which nonlocal terms of the action should form a correction to the Einstein Lagrangian arising via the replacement
$R\Rightarrow R+R^{\mu\nu}F(\Box)G_{\mu\nu}$. The nonlocal form factor of this correction $F(\Box)$ should be small in the GR domain, but it must considerably modify dynamics at the DE scale. Motivated by customary spectral representations for nonlocal quantities like
    \begin{eqnarray}
    F(\Box)=\int dm^2\,\frac{\alpha(m^2)}{m^2-\Box}
    \end{eqnarray}
we might try the following ansatz, $F(\Box)=\alpha/(m^2-\Box)$, corresponding to the situation when the spectral density $\alpha(m^2)$ is sharply peaked around some $m^2$ (cf. a similar discussion in \cite{DvaliHofmannKhoury}). As we will see, for $m^2\neq 0$ this immediately leads to a serious difficulty. Schematically the inverse propagator of the theory -- the kernel of the quadratic part of the action in metric perturbations $h_{\mu\nu}$ -- becomes
    \begin{eqnarray}
    -\Box+\alpha\frac{\Box^2}{m^2-\Box},
    \end{eqnarray}
where the squared d'Alembertian $\Box^2$ follows from four derivatives contained in the term bilinear in curvatures.
Then its physical modes are given by the two roots of this expression -- the solutions of the corresponding quadratic equation $\Box=m_\pm^2$. In addition to the massless graviton with $m_-^2=0$ massive modes with $m_+^2=O(m^2)$ appear and contribute a set of ghosts which cannot be eradicated by gauge transformations (for the latter have to be expended on cancelation of ghosts in the massless sector -- longitudinal and trace components of $h_{\mu\nu}$ subject to $\Box h_{\mu\nu}=0$.).

Therefore, only the case of $m^2=0$ remains, and as a first step to the nonlocal gravity we will consider the action
    \begin{eqnarray}
    S=
    \frac{M^2}2\int dx\,g^{1/2}\,\left\{\,-R+
    \alpha\,R^{\mu\nu}
    \frac1\Box\,G_{\mu\nu}\,\right\}            \label{action0}
    \end{eqnarray}
(for brevity we omit the surface integral that should accompany the Einstein Ricci scalar term). On the flat-space background this theory differs little from GR provided the dimensionless parameter $\alpha$ is small, $|\alpha|\ll 1$. Upper bound on $|\alpha|$ should follow from post-Newtonian corrections in this model. The additional effect of $\alpha$ is a small renormalization of the effective Planck mass. In the linearized theory we have an obvious relation
    \begin{eqnarray}
    S=-\frac{M^2(1-\alpha)}2\int dx\,g^{1/2}R+\alpha \,O[\,h_{\mu\nu}^3].
    \end{eqnarray}
which allows one to relate the constant $M$ to $M_P$ by Eq.(\ref{M_Prenorm}).

\section{Treatment of nonlocality: Schwinger-Keldysh technique vs Euclidean field theory}

At this point we have to address the treatment of nonlocality in (\ref{action0}) and (\ref{action}). In principle, handling the theories having a nonlocal action at the fundamental level is a sophisticated and very often an open issue, because their nonlocal equations of motion demand special care in setting their boundary value problem. Contrary to local field theories subject to a clear Cauchy  problem setup and local canonical commutation relations, nonlocal theories can have very ambiguous rules which are critical for physical predictions. In particular, the action (\ref{action0}) above requires specification of boundary conditions for the nonlocal Green's function $1/\Box$ which will necessarily violate causality in variational equations of motion for this action. Indeed, the action (\ref{action0}) effectively symmetrizes the kernel of the Green's function $G(x,y)$ of $1/\Box$, so that nonlocal terms in equations of motion
    \begin{equation}
    \frac{\delta S}{\delta g_{\mu\nu}(x)}\propto
    \nabla\nabla\int dy\,\big[\,G(x,y)+G(y,x)\,\Big]\,R(y)+...
    \end{equation}
($R(y)$ denoting a collection of curvatures) never have retarded nature even when $G(x,y)$ is the retarded propagator or satisfies any other type of boundary conditions \cite{DeffWood}. Therefore, these equations break causality because the behavior of the field at the point $x$ is not independent of the field values at the points $y$ belonging to the future light cone of $x$, $y^0>x^0$.

To avoid these ambiguities and potential inconsistencies we will once and for all assume that our nonlocal action is not fundamental. Rather it is the quantum effective action -- the generating functional of one-particle irreducible diagrams -- whose functional argument is the mean quantum field. This functional is necessarily nonlocal, and its nonlocality originates from quantum effects (by various mechanisms widely discussed in literature including \cite{quantum0,quantum1}). In this case boundary conditions for nonlocal operations are uniquely fixed by the choice of the initial (and/or final) quantum state, and manifest breakdown of causality in variational equations for this action is harmless under a proper treatment of their nonlocal terms.

To begin with, this causality breakdown does not immediately signify inconsistency in the calculation of scattering amplitudes or $\langle\,{\rm in}\,|\,{\rm out}\,\rangle$ matrix elements, because these amplitudes are determined by Feynman diagrammatic technique and do not have manifest retardation properties because they are not directly physically observable. Physically observable quantities like probabilities are bilinear combinations of scattering amplitudes and can always be represented as expectation values $\langle\,{\rm in}\,|\,\hat{\cal O}\,|\,{\rm in}\,\rangle$ of certain quantum operators $\hat{\cal O}$ in the initial quantum state $|\,{\rm in}\,\rangle$. For example, the probability of transition from this state to some final state $|\,{\rm fin}\,\rangle$
    \begin{eqnarray}
    P_{\,\rm in \to fin}=\langle\,{\rm in}\,|\,{\rm fin}\,\rangle\langle{\,\rm fin}\,|\,{\rm in}\,\rangle=
    \langle\,{\rm in}\,|\,\hat P_{\,\rm fin}|\,{\rm in}\,\rangle
    \end{eqnarray}
is an expectation value of the projector $\hat P_{\,\rm fin}\equiv|\,{\rm fin}\,\rangle\langle{\,\rm fin}\,|$  onto this final state. In contrast to in-out matrix elements these expectation values are subject to Schwinger-Keldysh diagrammatic technique \cite{SchwKeld,SchwKeld2} which guarantees causality of $\langle\,{\rm in}\,|\,\hat{\cal O}(x)\,|\,{\rm in}\,\rangle$. This property can be formulated as a retarded response of this average to the variation of the classical external source $J(y)$ coupled to the quantum fields in terms of which the operator $\hat{\cal O}(x)$ is built,
    \begin{eqnarray}
    \frac{\delta\langle\,{\rm in}\,|\,\hat{\cal O}(x)\,|\,{\rm in}\,\rangle}{\delta J(y)}=0,\quad x^0<y^0.
    \end{eqnarray}

This property is again not manifest and turns out to be the consequence of locality and unitarity of the original fundamental field theory (achieved via a complex set of cancellations between nonlocal terms with chronological and anti-chronological boundary conditions). However, there exists a class of problems for which a retarded nature of effective equations of motion explicitly follows from their quantum effective action calculated in Euclidean spacetime \cite{beyond}. This is a statement based on Schwinger-Keldysh technique \cite{SchwKeld} that for an appropriately defined initial quantum state $|{\rm in}\rangle$ the effective equations for the mean field
    \begin{eqnarray}
    g_{\mu\nu}=\langle{\,\rm in}\,|\,\hat g_{\mu\nu}|\,{\rm in}\,\rangle
    \end{eqnarray}
originate from the {\em Euclidean} quantum effective action $S=S_{\rm Euclidean}[g_{\mu\nu}]$ by the following procedure \cite{beyond}\footnote{We formulate this statement directly for the case of gravity theory with the expectation value of the metric field operator $\hat g_{\mu\nu}(x)$, though it is valid in a much wider context of a generic local field theory \cite{beyond}.}. Calculate nonlocal $S_{\rm Euclidean}[g_{\mu\nu}]$ and its variational derivative. In the Euclidean signature spacetime nonlocal quantities, relevant Green's functions and their variations are generally uniquely determined by their trivial (zero) boundary conditions at infinity, so that this variational derivative is unambiguous in Euclidean theory. Then make a transition to the Lorentzian signature and impose the {\em retarded} boundary conditions on the resulting nonlocal operators,
    \begin{eqnarray}
    \left.\frac{\delta S_{\rm Euclidean}}{\delta g_{\mu\nu}}\right|_{\;++++\,\;
    \Rightarrow\;-+++}^{\;\rm retarded}=0.   \label{EuclidLorentz}
    \end{eqnarray}
These equations are causal ($g_{\mu\nu}(x)$ depending only on the field behavior in the past of the point $x$) and satisfy all local gauge and diffeomorphism symmetries encoded in the original $S_{\rm Euclidean}[g_{\mu\nu}]$.

A similar treatment of a nonlocal action in \cite{DeffWood} was very reservedly called the "integration by parts trick" needing justification from the Schwinger-Keldysh technique. However, this technique only provides the causality of effective equations, but does not guarantee the Euclidean-Lorentzian relation (\ref{EuclidLorentz}). The latter is based, among other things, on the choice of the $|{\rm in}\rangle$-state.

We will assume that our model falls into the range of validity of this procedure, which implies a particular vacuum state $|{\rm in}\rangle$ and the one-loop approximation (in which it was proven to the first order of perturbation theory in \cite{Hartle-Horowitz} and to all orders in \cite{beyond}). The extension of this range is likely to include multi-loop orders and the $|{\rm in}\rangle$-state on the (A)dS background considered below, for which this state apparently coincides with the Euclidean Bunch-Davies vacuum.

\subsection{Asymptotically-flat space vs cosmological setup}
This subsection demonstrates trial application of the model (\ref{action0}) in cosmological setup with the purpose of generating DE. Though it will fail due to inconsistent treatment of boundary conditions, it is instructive to pass this exercise to see the importance of their careful treatment.

Like in papers on $f(R/\Box)$-gravity (see \cite{Odintsovetal} and references therein) stemming from \cite{nonloccosm}, but in contrast to \cite{nonloccosm} disregarding consistent treatment of boundary conditions for nonlocal operations, one can localize the nonlocal part of (\ref{action0})  with the aid of an auxiliary tensor field $\varphi^{\mu\nu}$. Then, the theory is equivalently described by the action
    \begin{eqnarray}
    &&S[\,g,\varphi\,]=
    \frac{M^2}2\int dx\,g^{1/2}\,\Big\{-R
    -2\alpha\,\varphi^{\mu\nu}R_{\mu\nu}\nonumber\\
    &&\qquad\qquad\qquad
    -\alpha\,\Big(\varphi^{\mu\nu}
    -\frac12\,g^{\mu\nu}\varphi\Big)\,
    \Box\,\varphi_{\mu\nu}\Big\}            \label{varphiaction}
    \end{eqnarray}
generating for $\varphi^{\mu\nu}$ the equation of motion  $\Box\varphi^{\mu\nu}=-G^{\mu\nu}$. Since (\ref{action0}) is understood as the Euclidean action with zero boundary conditions for $1/\Box$ at infinity, the auxiliary tensor field should satisfy the same Dirichlet boundary conditions $\varphi^{\mu\nu}|_{\,\infty}=0$,
and this is critically important for stability of the theory. Indeed, the field $\varphi^{\mu\nu}$ formally contains ghosts, but they do not indicate physical instability because they never exist as a free fields in the external lines of Feynman graphs. In the Lorentzian context of (\ref{EuclidLorentz}) this means that $\varphi^{\mu\nu}$ is given by a retarded solution, $\varphi^{\mu\nu}=-(1/\Box)_{\rm ret}G^{\mu\nu}$, and does not include free waves coming from asymptotic infinity.

Artificial nature of these ghosts is analogous to the case of the simplest ghost-free action that can be formally rendered nonlocal
    \begin{equation}
    S[\,\varphi\,]\equiv
    -\int dx\,\varphi \Box\varphi=
    -\int dx\,(\Box\varphi)
    \frac1{\Box}(\Box\varphi)    \nonumber
    \end{equation}
and further localized in terms of the auxiliary field $\psi$ with the action
    \begin{equation}
    S[\,\varphi,\psi\,]=
    \int dx\,\left(2\psi\,\Box\varphi
    +\psi\,\Box\psi\right).           \nonumber
    \end{equation}
This action is equivalent to the original one when $\psi$ is integrated out with the boundary conditions $(\psi+\varphi)\,|_{\,\infty}=0$.
After diagonalization this action features the ghost field $g\equiv\psi+\varphi$,
    \begin{eqnarray}
    S[\,\varphi,\psi\,]=
    \int dx\, \left(g\,\Box g
    -\varphi\,\Box\varphi\right).
    \end{eqnarray}
This ghost is however harmless because under the boundary conditions of the above type it identically vanishes in view of its equation of motion $\Box g=0$. In the presence of interaction, a nonvanishing $g$ exists in the intermediate states, but never arises in the asymptotic states, or external lines of Feynman graphs.

Main lesson to be drawn from the above example is that the actual particle content of the theory should be determined in terms of the original set of fields, whereas nonlocal reparameterizations can lead to artificial ghost modes which are actually eliminated by correct boundary conditions.\footnote{This, in particular, means that the ghost avoidance criteria derived in \cite{Odintsovetal} are not precise, because they are in fact applied to local scalar-tensor theories rather than to nonlocal ones. We have to reiterate here that a naive analysis of kinetic terms of auxiliary fields, which are usually used to localize a nonlocal action of the theory, cannot excusively serve as a criterion of the elimination of ghosts -- boundary conditions are equally important for that.} In our case this is the original formulation (\ref{action0}) in terms of the metric field $g_{\mu\nu}$. It indeed turns out to be ghost-free on the flat-space background, because the quadratic part of the action coincides with the Einstein one.\footnote{A similar mechanism of eliminating ghosts by boundary conditions was recently used in \cite{Maldacena}. However, in contrast to our model the ghost modes of the conformal gravity in \cite{Maldacena} are higher-derivative ones and are essentially nonlinear. Therefore, the non-ghost nature of the theory requires further verification even after integrating these ghosts out .}

In the local representation (\ref{varphiaction}) our model could be directly applied to the FRW cosmology for the purpose of finding the accelerated stage of cosmic expansion. It is easy to find a (quasi) de Sitter point of the cosmological evolution. Indeed, with the natural Lorentz-invariant ansatz for the auxiliary field $\varphi^{\mu\nu}\simeq \frac14\Phi\,g^{\mu\nu}$, which is supposed to be valid close to a certain moment $t_0$ corresponding to the present epoch, the cosmological evolution for the action (\ref{varphiaction}) can be compatible with the current DE data. By an appropriate choice of initial conditions the Hubble factor $H=\dot a/a$, the field $\Phi$ and the parameter of the effective equation of state $w=-1-2\dot H/3H^2$ can satisfy at $t_0$ the following relations
    \begin{eqnarray}
    &&\dot\Phi_0=-4\,\frac{H_0}\sigma,\nonumber\\
    &&w_0=-1,\quad\dot w_0=-16\,H_0\frac{2\sigma-1}{3\sigma^2+2}
    \end{eqnarray}
Here the quantity $\sigma=(2\alpha/3)^{1/2}(2+\alpha\Phi_0-3\alpha)^{-1/2}$ is determined by the value of the field $\Phi_0=\Phi(t_0)$.
If it is chosen to satisfy $\sigma=O(1)>1/2$ we have $\dot w_0=O(1)\times H_0<0$ which yields a pace of change in the effective equation of state compatible with the horizon scale.

These preliminary estimates could have served as a starting point for a quantitative comparison with the DE scenario. However, a formal application of (\ref{action0}) to the FRW setup disregards nontrivial boundary conditions in cosmology. To see this, note that fine tuning initial conditions for $\Phi$ to the DE data would generally contradict zero boundary conditions for the auxiliary tensor field $\varphi^{\mu\nu}\propto \Phi g^{\mu\nu}$, not to mention that the cosmological FRW setup does not in principle match with the asymptotically-flat framework of the action (\ref{action0}). Therefore we have to extrapolate the definition (\ref{action0}) to nontrivial backgrounds including, first of all, the de Sitter spacetime and change our technique -- instead of localization method with an auxiliary tensor field work directly in the original metric representation. This will help us to look for the (A)dS solution in the covariant language circumventing explicit FRW metric ansatz.

This approach to the action (\ref{action0}) suffers from a serious difficulty. Ricci curvature for the (A)dS background (\ref{Ricci}) is covariantly constant, and the nonlocal part of (\ref{action0}) turns out to be infrared divergent, $(1/\Box)g_{\mu\nu}=\infty$ -- the property that can hardly be cured by some choice of boundary conditions for $1/\Box$. This means that the action (\ref{action0}) should be modified to regulate this type of divergences.

\section{Stable (A)dS background}
We will regulate the action (\ref{action0}) by adding to the covariant d'Alembertian the matrix-valued potential term built of a generic combination of tensor structures linear in the curvature. This brings us to the six-parameter family of nonlocal action functionals (\ref{action})-(\ref{potential}) introduced in Sect.II. Of course, such a modification of the original action (\ref{action0}) leaves its linear approximation on a flat background intact, because it deals with $O[h_{\mu\nu}^3]$-terms, whereas its dimensionless parameters will be restricted by the requirement of a stable dS or AdS solution in the model.

The action of the matrix valued  potential $\hat P$ on $g_{\mu\nu}$ is given by the relation
    \begin{eqnarray}
    &&\hat P\,g_{\mu\nu}\equiv P_{\mu\nu}^{\;\;\;\;\alpha\beta}\,g_{\alpha\beta}
    =A\,R_{\mu\nu}
    +B\,R\,g_{\mu\nu},      \label{Pg}
    \end{eqnarray}
where the coefficients $A$ and $B$ are defined by Eqs.(\ref{A})-(\ref{B}), so that the Green's function  $1/(\Box+\hat P)$ acting on Ricci and Einstein tensors in (\ref{action}) is well defined even for the maximally symmetric (A)dS background with the covariantly constant curvatures (\ref{Riemann})-(\ref{Ricci}). In this case the above relation simplifies to the equation
    \begin{eqnarray}
    &&\hat P\,g_{\mu\nu}
    =(A+4B)\Lambda g_{\mu\nu}    \label{Pg1}
    \end{eqnarray}
which has two obvious corollaries
    \begin{eqnarray}
    &&\left.\frac1{\Box+\hat P}\,
    R_{\mu\nu}\right|_{\,\rm (A)dS}=
    \Lambda\hat P^{-1}g_{\mu\nu}=
    \frac1{A+4B}\,g_{\mu\nu},      \label{GreenfunconR}\\
    &&\left.\frac1{\Box+\hat P}\,
    G_{\mu\nu}\right|_{\,\rm (A)dS}=
    -\frac1{A+4B}\,g_{\mu\nu}.     \label{GreenfunconG}
    \end{eqnarray}
These corollaries follow from the fact that on a maximally symmetric background $\Box$ commutes with $\hat P$ and annihilates $R_{\mu\nu}=\Lambda g_{\mu\nu}$.

Let us prove now that under the relation (\ref{relation}) between the parameters of the potential (\ref{potential}) the equation of motion for (\ref{action}) has the (A)dS solution with an {\em arbitrary} value of the cosmological constant $\Lambda$. The simplest method to see this is to apply to the action the conformal metric variation (more cumbersome straightforward calculation will be presented in Appendix).

For this introduce the local conformal variation with the parameter $\delta\sigma=\delta\sigma(x)$,
    \begin{eqnarray}
    \delta_\sigma=\int d^4x\,\delta\sigma\,g_{\alpha\beta}\,\frac\delta{\delta g_{\alpha\beta}}.
    \end{eqnarray}
Under its action various quantities in (\ref{action}) transform according to their conformal weights,
    \begin{eqnarray}
    &&\delta_\sigma g_{\mu\nu}=\delta\sigma\,g_{\mu\nu},\quad \delta_\sigma g^{1/2}=2\,g^{1/2}\delta\sigma,\nonumber\\
    &&\delta_\sigma R_{\mu\nu}=O(\nabla),\,\,\,\,\delta_\sigma R=-\delta\sigma\, R+O(\nabla), \nonumber\\
    &&\delta_\sigma R^{\mu\nu}=
    -2\delta\sigma\, R^{\mu\nu}+O(\nabla),\,\,\,\,
    \delta_\sigma\hat P=-\delta\sigma\,\hat P+O(\nabla), \nonumber\\
    &&\delta_\sigma\Box=O(\nabla),
    \end{eqnarray}
modulo the derivatives $O(\nabla)$ acting on $\delta\sigma(x)$ and these quantities themselves. Then the conformal variation of (\ref{action}) {\em on the (A)dS background} reads
    \begin{eqnarray}
    &&\delta_\sigma S=
    \frac{M^2}2 \int d^4x\,g^{1/2}\left\{-R+\alpha\,R^{\alpha\beta}\frac1{\Box+\hat P}\,G_{\alpha\beta}\,\right\}\delta \sigma\nonumber\\
    &&\quad\quad
    =-2M^2\Lambda\left(1+\frac\alpha{A+4B}\right)\int d^4x\,g^{1/2}\delta\sigma,
    \end{eqnarray}
so that the corresponding variational derivative equals
    \begin{eqnarray}
    &&\left.\frac{\delta S}{\delta\sigma}\,\right|_{\;\rm (A)dS}=
    -2M^2\Lambda
    \left(1+\frac\alpha{A+4B}\right)\,g^{1/2}.   \label{500}
    \end{eqnarray}
Since all tensor quantities on this background algebraically express via $g_{\mu\nu}$ the metric variational derivative of the action reduces to this conformal variation. Then the resulting equation of motion
    \begin{eqnarray}
    \left.\frac{\delta S}{\delta g_{\mu\nu}}\,\right|_{\;\rm (A)dS}=\left.\frac14\,g^{\mu\nu}\frac{\delta S}{\delta\sigma}\,\right|_{\;\rm (A)dS}=0    \label{501}
    \end{eqnarray}
holds with an arbitrary value of the effective cosmological constant $\Lambda$ when the parameters in Eqs.(\ref{action})-(\ref{potential}) satisfy the relation (\ref{relation}).

Note that the existence of the (A)dS solution with an arbitrary $\Lambda$ is neither the result of the local Weyl invariance of the theory, nor even its global scale invariance. Rather this is the corollary of the relation (\ref{relation}) which, in virtue of Eq.(\ref{GreenfunconG}), guarantees the vanishing on-shell value of the action,
    \begin{eqnarray}
    S\,\big|_{\,\rm (A)dS}=0    \label{zeroonshell}
    \end{eqnarray}
Thus, this solution is another vacuum -- a direct analogue of the flat-space one.

Another remarkable corollary of Eq.(\ref{relation}) is that the stability of the (A)dS solution against ghost and tachyon excitations is guaranteed by only one additional restriction (\ref{Crelation}) on the parameter $C$ defined by (\ref{C}). In principle, the hope to eliminate ghosts and tachyons from the quadratic part of the action $S_{(2)}$ on the (A)dS background is based on the following observation.

In the DeWitt gauge on metric perturbations (\ref{DWgauge}) $S_{(2)}$ contains only two structures $h^{\mu\nu}\!\times h_{\mu\nu}$ and $h\!\times h$, or equivalently $\bar h^{\mu\nu}\times \bar h_{\mu\nu}$ and $h\times h$, where $\bar h_{\mu\nu}$ is a traceless part of $h_{\mu\nu}$. Since the potential term $\hat P$ on the maximally symmetric background commutes with $\Box$ and equals the linear combination (\ref{PonAdS}) of projectors on traceless and trace subspaces, the nonlocal parts of these two structures have the form
    \begin{eqnarray}
    \bar h^{\mu\nu}\frac1{\Box-C\Lambda}\,\bar h_{\mu\nu},\quad
    h\,\frac1{\Box-\alpha\Lambda}\,h
    \end{eqnarray}
where the Green's function in the trace sector follows from the equation
    \begin{eqnarray}
    (\Box+\hat P)\,g_{\mu\nu}h=g_{\mu\nu}(\Box-\alpha\Lambda)\,h
    \end{eqnarray}
which is also based on (\ref{relation}) and (\ref{Pg1}). As in the flat-space background discussion above, the ghosts necessarily appear if these nonlocalities are nonvanishing in $S_{(2)}$, because the dispersion equation for $\Box$ becomes quadratic and generates a doubled set of physical modes with $\Box=m_\pm^2$ (with $C\Lambda$ or $\alpha\Lambda$ playing the role of a nonvanishing $m^2$). Therefore it seems a priori possible to cancel these two nonlocal structures and provide the right signs of the remaining local terms by the appropriate choice of the five free parameters in (\ref{potential}).

Curious fact is that in the DeWitt gauge (\ref{DWgauge}) $S_{(2)}$ has a very simple form
    \begin{eqnarray}
    &&S_{(2)}=\frac{M^2_{\rm eff}}2
    \int d^4x\,g^{1/2}\left\{-\frac14\,h^{\mu\nu}\Box\, h_{\mu\nu}+\frac1{16}\,h\,\Box\,h\right.
    \nonumber\\
    &&\qquad\qquad-\frac14\,\left(C-\frac43\right)\,
    \Lambda\,h_{\mu\nu}^2
    +\frac1{16}\left(C-\frac43\right)\,
    \Lambda\,h^2\nonumber\\
    &&\qquad\qquad
    -\frac{\Lambda^2}4\,\left(C-\frac23\right)^2\,
    \left(h^{\mu\nu}\frac1{\Box+\hat P}\,h_{\mu\nu}\right.\nonumber\\
    &&\qquad\qquad\qquad\qquad\qquad\quad
    \left.\left.-\frac14\,h\,
    \frac1{\Box-\alpha\Lambda}\,h\right)\,\right\}, \label{s_2}
    \end{eqnarray}
where the effective Planck mass $M_{\rm eff}$ is given by Eq.(\ref{effectivemass}).

A rather lengthy technical derivation of this result is presented in Appendix. It is strongly based on the relation (\ref{relation}), Eqs. (\ref{GreenfunconR})-(\ref{GreenfunconG}) and, what is less pronounced but equally important, on a symmetric variational relation for the Green's function (\ref{symvar}) and the possibility of integrating by parts without extra surface terms. These properties are guaranteed by the Euclidean spacetime signature in the action (\ref{action}) and regularity of Green's functions on a closed compact $S^4$ for the de Sitter case or their boundary conditions at the AdS boundary for the Anti-de Sitter case.

Again, it is worth mentioning here that this rule of free integration by parts, that was used throughout the series of papers \cite{DeffWood,nonloccosm,Woodard2} and interpreted there as a trick, makes sense only within the relation between the Schwinger-Keldysh technique and Euclidean QFT and applies only in the Euclidean spacetime. Below we will see that, once the effective equations in the physical Lorentzian spacetime have been obtained by Eq.(\ref{EuclidLorentz}), indiscriminate omission of surface terms under integration by parts becomes illegitimate and leads to wrong properties of free propagating modes of $h_{\mu\nu}$.

Remarkably, the expression (\ref{s_2}) depends only on the traceless part (\ref{traceless}) of the metric perturbation, because the combination of tensor contractions $h^{\mu\nu}\times\, h_{\mu\nu}-\frac14\,h\times h$ equals $\bar h^{\mu\nu}\times\,\bar h_{\mu\nu}$. This property, apart from explicit calculation done in the Appendix, can be derived from the scaling transformation of the action (\ref{action})
    \begin{eqnarray}
    S[\,(1+\varepsilon)g_{\mu\nu}]
    =(1+\varepsilon)\,S[\,g_{\mu\nu}]     \label{scaling}
    \end{eqnarray}
under a global dilatation of the metric, $\nabla_\mu\varepsilon=0$. This implies that the quadratic part of the action (in fact on an arbitrary background) identically vanishes for global conformal perturbations $h_{\mu\nu}=\varepsilon\,g_{\mu\nu}$ obviously satisfying the DeWitt gauge (\ref{DWgauge})
    \begin{eqnarray}
    S_{(2)}\big|_{\;h_{\mu\nu}
    =\varepsilon\,g_{\mu\nu}}=0,\quad \varepsilon={\rm const}.
    \end{eqnarray}
This observation serves as an independent check of explicit calculations of $S_{(2)}$ in Appendix and uniquely determines the minus one quarter ratio of the coefficients of $h\times\,h$ and $h^{\mu\nu}\times\, h_{\mu\nu}$ structures (except their local kinetic terms $h\Box h$ and $h^{\mu\nu}\Box h_{\mu\nu}$ which identically vanish for $h_{\mu\nu}=\varepsilon\,g_{\mu\nu}$ with a constant $\varepsilon$).

Thus  in the DeWitt gauge the quadratic part of the action (\ref{s_2}) takes the form (\ref{quadrS}) quadratic in the traceless part of metric perturbations. Moreover, in view of the projector properties of the potential term (\ref{PonAdS}) it simplifies even further
    \begin{eqnarray}
    &&S_{(2)}=-\frac{M^2_{\rm eff}}8
    \int d^4x\,g^{1/2}\,
    \bar h^{\mu\nu}\frac{\left(\Box
    -\frac23\Lambda\right)^2}{\Box-C\Lambda}\,
     \bar h_{\mu\nu}.                         \label{quadrS1}
    \end{eqnarray}
With an arbitrary value of $C$ the characteristic equation for  physical modes still yields only one root $\Box=\frac23\Lambda$, but this is the second order root which for $C\neq\frac23$ corresponds to double poles in the propagator and signifies presence of higher-derivative ghosts.

Therefore, the requirement of absence of ghosts imposes only one extra equation (\ref{Crelation}) for $C$, $C=2/3$, and the positivity requirement for $M_{\rm eff}^2$. Bearing in mind that $|\alpha|\ll 1$, this selects two admissible intervals for the parameter $B$ in the case of a positive $\alpha$,
    \begin{eqnarray}
    B<-\frac\alpha2,\,\,\,\,\,B>0,       \label{positivealpha}
    \end{eqnarray}
and even more interesting compact range of this parameter for a negative $\alpha$,
    \begin{eqnarray}
    &&0<B<-\frac\alpha2,\,\,\,\,\,\alpha<0.   \label{negativealpha}
    \end{eqnarray}

\section{Free propagating modes and gravitational potentials in the (A)dS phase}
In order to obtain correct linearized equations of motion we need a quadratic part of the action without gauge-fixing. It is invariant under the linearized gauge transformations $h_{\mu\nu}\to h_{\mu\nu}+\Delta^f h_{\mu\nu}$ with the vector diffeomorphism parameters $f_\mu=g_{\mu\nu} f^\nu$,
    \begin{eqnarray}
    \Delta^f h_{\mu\nu}=
    \nabla_\mu f_\nu+\nabla_\nu f_\mu,  \label{gaugetransform}
    \end{eqnarray}
where covariant derivatives are determined with respect to the background metric. The invariant $S_{(2)}$ can be obtained from (\ref{s_2}) by representing $h^{\mu\nu}$ in the DeWitt gauge as the projection to this gauge of the non-gauged field,
    \begin{eqnarray}
    h_{\mu\nu}\Big|_{\,\chi^\alpha=0}
    =h_{\mu\nu}-2\,\nabla_{(\mu}
    \frac{\delta^\alpha_{\nu)}}{\Box+\Lambda}\,
    \chi_\alpha.                            \label{gaugedh}
    \end{eqnarray}
Here it is worth reminding the definition of the DeWitt gauge condition functions (\ref{DWgauge}) for which the vector field operator $(\Box+\Lambda)\,\delta^\mu_\nu$ plays the role of the Faddeev-Popov operator
    \begin{eqnarray}
    &&\chi_\alpha=g_{\alpha\mu}\chi^\mu,\quad
    \chi^\mu\equiv\nabla_\nu h^{\mu\nu}
    -\frac12\,\nabla^\mu h,         \label{DWgaugefunctions}\\
    &&\Delta^f\chi^\mu
    =(\Box+\Lambda)\,f^\mu.
    \end{eqnarray}

The result of substituting (\ref{gaugedh}) in Eq.(\ref{s_2}) for $C=\frac23$ reads
    \begin{eqnarray}
    &&S_{(2)}=\frac{M_{\rm eff}^2}2\int d^4x\,g^{1/2}\left\{\,\frac14\,h^{\mu\nu}
    \left(-\Box+\frac23\,\Lambda\right)\, h_{\mu\nu}
    \right.\nonumber\\
    &&\qquad\qquad\qquad\qquad\quad
    -\frac18\,h\left(-\Box-\frac23\,\Lambda\right)\,h
    -\frac12\,\chi_\mu^2\nonumber\\
    &&\qquad\qquad\qquad\qquad\quad\left.
    -\frac1{4}\,R_{(1)}\,
    \frac1{\Box+2\Lambda}\,R_{(1)}\right\},   \label{quadrS2inv}
    \end{eqnarray}
where $R_{(1)}$ is the linearized Ricci scalar on the (A)dS background which has a form of the combination of two terms linear respectively in $\nabla_\mu\chi^\mu$ and $h$,
    \begin{eqnarray}
    &&R_{(1)}\equiv\nabla^\mu\nabla^\nu h_{\mu\nu}
    -\Box h-\Lambda\,h\nonumber\\
    &&\qquad\qquad\qquad
    =\nabla_\mu\chi^\mu
    -\frac12\,(\Box+2\Lambda)\,h.            \label{R1}
    \end{eqnarray}

Interestingly, the first two lines of $S_{(2)}$ above coincide with the quadratic part on the (A)dS background of the Einstein-Hilbert action with the $\Lambda$ term \cite{AllenTseytlinAdS}. Therefore, this part of $S_{(2)}$ is invariant under (\ref{gaugetransform}), while the invariance of the last term of (\ref{quadrS2inv}) directly follows from the invariance of $R_{(1)}$. The form of (\ref{quadrS2inv}) shows that despite a complicated tensor structure of nonlocal term in (\ref{action}) near the (A)dS background it reduces to the Ricci scalar sector similar to the $R\frac1\Box R$ distortion of Einstein theory considered in \cite{nonloccosm,DeffWood,Woodard2}.

Now, according to the relation (\ref{EuclidLorentz}) between Euclidean QFT and causal effective equations for mean field we calculate the variational derivative of $S_{(2)}$ with respect to $h_{\mu\nu}$, go over to the Lorentzian metric signature and establish retardation of all nonlocal operations. This gives the following equation for free propagating modes of the {\em mean (expectation value)} metric field $h_{\mu\nu}=\langle{\,\rm in}\,|\,\hat h_{\mu\nu}|\,{\rm in}\,\rangle$,
    \begin{eqnarray}
    &&\!\!\!\!\!\!\!\!\left.\frac4{M_{\rm eff}^2}g^{-1/2}\frac{\delta S_{(2)}}{\delta h^{\mu\nu}}\right|_{\;++++\,\;
    \Rightarrow\;-+++}^{\;\rm retarded}\nonumber\\
    &&\nonumber\\
    &&\quad
    =\left(-\Box+\frac23\,\Lambda\right) h_{\mu\nu}
    +\frac12\,g_{\mu\nu}\left(\Box+\frac23\,\Lambda\right)h
    \nonumber\\
    &&\qquad\quad+\frac12\, g_{\mu\nu}R_{(1)}
    +2\,\nabla_{(\mu}\varPhi_{\nu)}
    -g_{\mu\nu}\nabla_\alpha\varPhi^\alpha=0,  \label{freequation}
    \end{eqnarray}
where $\varPhi^\mu$ is the nonlocal function
    \begin{eqnarray}
    &&\varPhi^\mu=\chi^\mu-\frac12\,\nabla^\mu
    \frac1{\Box+2\Lambda}\,
    \Big|_{\;\rm ret} R_{(1)}          \label{varPhi}
    \end{eqnarray}
whose nonlocality is given by the retarded Green's function .

The trace of this effective equation gives
    \begin{eqnarray}
    &&\frac4{M_{\rm eff}^2}g^{-1/2}g^{\mu\nu}\frac{\delta S_{(2)}}{\delta h^{\mu\nu}}=2\,\nabla_\mu(\chi^\mu-\varPhi^\mu)\nonumber\\
    &&\qquad\qquad\quad=\Box\left[\,\frac1{\Box+2\Lambda}\,\right]_{\;\rm ret}\!R_{(1)}\nonumber\\
    &&\qquad\qquad\quad=R_{(1)}
    -\frac{2\Lambda}{\Box+2\Lambda}\,
    \Big|_{\;\rm ret} R_{(1)}=0.     \label{R1equation}
    \end{eqnarray}
The last equation  yields not only the homogeneous differential equation for $R_{(1)}$ but also its zero initial conditions at past infinity because of the retarded nature of the Green's function. Acting on this equation by the operator $\Box+2\Lambda$ we immediately get the initial value problem,
    \begin{eqnarray}
    &&\Box R_{(1)}=0,\\
    &&R_{(1)}\big|_{\;x^0\to-\infty}=0,\quad
    \partial_0 R_{(1)}\big|_{\;x^0\to-\infty}=0,
    \end{eqnarray}
with the identically vanishing solution. Therefore the linearized Ricci scalar of the free propagating wave is vanishing throughout the entire (A)dS spacetime
    \begin{eqnarray}
    R_{(1)}(x)=0.     \label{R1equation1}
    \end{eqnarray}
As a result the nonlocal function (\ref{varPhi}) coincides with the local DeWitt gauge condition function, $\varPhi^\mu=\chi^\mu$, and Eq.(\ref{freequation}) becomes absolutely identical to the linearized Einstein equations on the (A)dS background.

Now it is high time to impose the DeWitt gauge (\ref{DWgauge}) in which Eq.(\ref{R1equation1}) reduces to
    \begin{eqnarray}
    (\Box+2\Lambda)h=0.     \label{h_equation}
    \end{eqnarray}
Similarly to the Feynman gauge in electrodynamics, in this relativistic gauge all components of $h_{\mu\nu}$ are propagating, but their gauge ambiguity is not completely fixed and admits residual gauge transformations (\ref{gaugetransform}) with the parameter $f_\mu$ satisfying the equation
    \begin{eqnarray}
    (\Box+\Lambda)f_\mu=0.    \label{residualdiffequation}
    \end{eqnarray}
By the usual procedure these transformations can be used to select two polarizations -- free physical modes $h_{\mu\nu}^{\rm phys}=h_{\mu\nu}+\nabla_\mu f_\nu+\nabla_\nu f_\mu$. In particular, they can nullify boundary conditions for $h$ on any initial Cauchy surface $\Sigma$, so that this trace identically vanishes in view of the homogeneous equation (\ref{h_equation}) and makes the physical modes transverse and traceless as in the Einstein theory with a $\Lambda$-term,
    \begin{eqnarray}
    \nabla^\nu h_{\mu\nu}^{\rm phys}=0,\quad
    h^{\rm phys}=0.
    \end{eqnarray}
Indeed, under these transformations $h^{\rm phys}=h+2\nabla_\mu f^\mu$, and both $\nabla_\mu f^\mu|_{\,\Sigma}$ and $\partial_0\nabla_\mu f^\mu|_{\,\Sigma}$ can be chosen to provide zero initial data for $h^{\rm phys}$ on $\Sigma$. The remaining three pairs of initial data for $f^\mu$ accomplishes the counting of the physical degrees of freedom among spatial components of $h_{\mu\nu}$, $6-1-3=2$, while the four lapse and shift functions $h_{0\mu}$, as usual, express via the constraint equations of motion $\delta S_{(2)}/\delta h_{0\mu}=0$.\footnote{Or equivalently, when they are treated as propagating modes subject to second order in time differential equations, their initial data express via $\chi^\mu|_{\,\Sigma}=0$ and $\partial_0\chi^\mu|_{\,\Sigma}=0$.}

It is important that a priori in the DeWitt gauge the equation of motion (\ref{freequation}) is not local even though the function (\ref{varPhi}) seems getting localized, cf. Eq.(\ref{R1}). This is because the seemingly correct equation
    \begin{eqnarray}
    \left[\,\frac1{\Box+2\Lambda}\,\right]_{\;\rm ret}\!(\overrightarrow{\Box}
    +2\Lambda)\,\varphi(x)=\varphi(x)  \label{retonbox}
    \end{eqnarray}
does not hold for the field $\varphi(x)$ with a noncompact support extending to past infinity. This is exactly the case of free propagating modes, when integration by parts of ${\Box}$ yields extra surface terms -- the situation different from the Euclidean QFT side of the relation (\ref{EuclidLorentz}), as it was mentioned in the previous section. Disregarding these terms would lead to inconsistent equations of motion breaking their diffeomorphism invariance.\footnote{Indiscriminate use of Eq.(\ref{retonbox}), which leads to the local equation (\ref{freequation}) in the DeWitt gauge, would imply the wrong trace equation $\Box h=0$ different from (\ref{h_equation}). This would contradict the corollary of Eq.(\ref{residualdiffequation}), $(\Box+2\Lambda)\nabla_\mu f^\mu=0$, because $h$ and $h^{\rm phys}$ differing by $2\nabla_\mu f^\mu$ would not satisfy one and the same equation as they should.}

Opposite situation occurs when we consider retarded gravitational potentials $h_{\mu\nu}$ generated by matter sources with a stress tensor $T_{\mu\nu}$ having a compact support. Then $h_{\mu\nu}$ also has a compact support in the past, and Eq.(\ref{retonbox}) applies to the calculation of the vector function (\ref{varPhi}). In the DeWitt gauge it becomes local, $\varPhi^\mu|_{\,\chi^\alpha=0}=\frac14\nabla^\mu h$, and the equation (\ref{freequation}) for the retarded potential also takes the local form,
    \begin{equation}
    \left(-\Box+\frac23\Lambda\right) h_{\mu\nu}
    +\frac12\nabla_\mu\nabla_\nu h
    -\frac\Lambda6 g_{\mu\nu}h
    =\frac2{M_{\rm eff}^2}T_{\mu\nu},         \label{coupledmode}
    \end{equation}
with matter stress tensor coupled to $h_{\mu\nu}$ via the effective Planck mass (\ref{effectivemass}).

The retarded solution of this equation for the trace part of $h_{\mu\nu}$ immediately reads
    \begin{eqnarray}
    M_{\rm eff}^2 h=-4\,\frac1\Box\,
    \Big|_{\;\rm ret}\!T.
    \end{eqnarray}
After careful commutation of covariant derivatives and the Green's function of the tensor operator $(-\Box+\frac23\,\Lambda)\delta_{\mu\nu}^{\;\;\;\alpha\beta}$,
    \begin{eqnarray}
    &&\frac{\delta_{\mu\nu}^{\;\;\;\alpha\beta}}
    {-\Box+\frac23\Lambda}\,
    \nabla_\alpha\nabla_\beta=-\nabla_\mu\nabla_\nu
    \frac1{\Box+2\Lambda}\nonumber\\
    &&\qquad\qquad+\frac{2\Lambda}3 g_{\mu\nu}\frac\Box{\big(\!-\Box+\frac23\Lambda\,\big)\,
    \big(\,\Box+2\Lambda\,\big)},
    \end{eqnarray}
one finally finds the gravitational potential of matter sources. In the {\em DeWitt gauge} it takes the form
    \begin{eqnarray}
    &&h_{\mu\nu}
    =\frac{16\pi G_{\rm eff}}{-\Box+\frac23\Lambda}\,\left[\,T_{\mu\nu}
    +g_{\mu\nu}\,
    \frac{\Box-2\Lambda}{\Box+2\Lambda}\,
    \frac{\Lambda}{3\Box}\,T\,\right]_{\;\rm ret}\!
    \nonumber\\
    &&\qquad\quad
    -\nabla_\mu\nabla_\nu \frac{16\pi G_{\rm eff}}{(\Box+2\Lambda)\,\Box}\,
    \Big|_{\;\rm ret}\!T.     \label{matpot}
    \end{eqnarray}
Here of course retardation prescription applies to all Green's functions, the last term represents a pure gauge transformation and $G_{\rm eff}\equiv 1/8\pi M_{\rm eff}^2$ is the effective gravitational constant vs the Newton one $G_N=1/8\pi M_P^2$,
    \begin{eqnarray}
    G_{\rm eff}=\frac{\alpha(1-\alpha)}
    {8B(2B+\alpha)}\,G_N.                 \label{Geff}
    \end{eqnarray}
This proves the gauge invariant part of this potential advocated in Sect.II.

The expression (\ref{matpot}) differs from the GR analog by the tensor structure -- the nonlocal combination in the first line of (\ref{matpot}) vs the GR structure $T_{\mu\nu}-\frac12g_{\mu\nu}T$. From the viewpoint of expansion in $|\alpha|\ll 1$ this implies for non-relativistic sources certain $O(1)$ corrections. What is much more interesting, it yields an unexpected bonus in the form of a possible dark matter simulation -- $O(1/|\alpha|)$ amplification of the gravitational attraction due to the replacement of the Newton gravitational constant $G_N$ by $G_{\rm eff}\sim G_N/|\alpha|$ with $|\alpha|\ll 1$. This necessarily happens in the case (\ref{negativealpha}) of a negative $\alpha$, because the factor $\alpha/8B(2B+\alpha)\geq 1/|\alpha|$ and
    \begin{eqnarray}
    G_{\rm eff}\geq\frac{1-\alpha}{|\alpha|}
    \,G_N\gg G_N.                                \label{10}
    \end{eqnarray}

For a positive $\alpha$ in the domain (\ref{positivealpha}) the theory is also strongly coupled with $G_{\rm eff}>G_N$ for the following two intervals of $B$,
    \begin{eqnarray}
    -\frac{\sqrt\alpha+\alpha}4
    <B<-\frac\alpha2,\quad
    0<B<\frac{\sqrt\alpha-\alpha}4,       \label{strongcoupling}
    \end{eqnarray}
and might correspond to the DM phenomenon. In the rest of this domain it is on the contrary weakly coupled with $G_{\rm eff}<G_N$. At the crossover between these two regions of strong and weak coupling, $|B|\simeq\sqrt\alpha/4$, both Newton and effective gravitational coupling constants can be of the same order of magnitude, $G_N/G_{\rm eff}=O(1)$ even for a very small $\alpha\ll 1$, which together with (\ref{10}) leaves a large window for a possible strength of DM attraction relative to the GR behavior.

\section{Conclusions}
Thus we have a class of generally covariant nonlocal gravity models which have a GR limit and also a stable de Sitter or Anti-de Sitter background with an arbitrary value of its cosmological constant. This background carries as free propagating modes two massless gravitons identical to those of Einstein theory with a cosmological term. These gravitons are coupled to matter with a variable effective gravitational coupling which can be stronger than the Newton coupling in general relativity theory. The interpretation of these models looks as follows.

The theory with the action (\ref{action}) has two phases. For short distances corresponding to the range of wavelengths with
    \begin{eqnarray}
    \nabla\nabla\sim\Box\gg R
    \end{eqnarray}
this is a GR phase on the zero curvature background with small $O(\alpha)\times R/\Box$ corrections of higher orders in spacetime curvature (collectively denoted by $R$). This regime might apply to galactic, Solar system and other small scale phenomena and is likely to pass all general relativistic tests for a sufficiently small $\alpha$. Disturbing property of perturbation theory in this phase could be the presence of poles $\sim(1/\Box)R$ in the vertices of (\ref{action}) which could contribute too strong to graviton scattering. But these contributions vanish on shell $R_{\mu\nu}=0$ provided, perhaps, that the Riemann term is forbidden in the potential (\ref{potential}) of the tensor differential operator which specifies a nonlocal part of (\ref{action}), $a=0$.

Another phase of the theory corresponds to the infrared wavelengths range
    \begin{eqnarray}
    \nabla\nabla\ll R
    \end{eqnarray}
in which a stable (A)dS background exists and the modified gravitational potential of matter sources is given by Eq.(\ref{matpot}). This equation is valid for the perturbation range $|\,\delta R^\mu_\nu\,|\sim|\,\nabla\nabla h^\mu_\nu\,|\ll \Lambda$ and $|\,h^\mu_\nu\,|\ll 1$, which is equivalent in virtue of Eq.(\ref{coupledmode}) to very small matter densities,
    \begin{eqnarray}
    |\,T^\mu_\nu\,|\sim M_{\rm eff}^2
    \Lambda\, |\,h^\mu_\nu\,|
    \ll M_{\rm eff}^2\Lambda,                        \label{smallT}
    \end{eqnarray}
characteristic of galaxy, galaxy cluster and horizon scales for which DE and DM modification of gravity theory becomes important. Thus, nonlocal gravity interpolates between GR theory and its infrared modification. The latter is likely to generate a stable ghost-free stage of cosmic acceleration and, perhaps, even simulate the DM effect on rotation curves in a strong coupling domain (\ref{negativealpha}) and  (\ref{strongcoupling}).

How realistic is this picture from the viewpoint of the concordance model of DE and DM? There are open problems which might derail its viability. The most serious objection is that absence of ghosts is guaranteed only on two backgrounds -- flat space and (A)dS ones. Outside of these two solutions the theory is most likely unstable, which perhaps can be interpreted as a kind of attraction mechanism pushing the system to one of its two stable phases. Therefore, inclusion of matter sources can be consistently done only within perturbation theory in the vicinity of these two solutions, that is when the total matter stress tensor $T_{\mu\nu}$ is treated as a perturbation. In fact, the range of energies (\ref{smallT}) is a condition justifying this perturbation theory on the (A)dS background. A similar range on the flat background is of course bounded by the Planck scale which is not at all restrictive, because we anyway stay in the infrared domain of effective gravity theory.

It goes without saying that the model should undergo tests on consistency of post-Newtonian corrections, the magnitude of the DM phenomenon should be compared with observations, the effect of nonlocal stress tensor trace term of (\ref{matpot}) should be studied, etc. Moreover, the mechanism should be found, by which the model picks up a concrete scale $|\,T^\mu_\nu\,|\sim M_{\rm eff}^2\Lambda$ of a crossover from the GR regime to cosmic acceleration -- necessary element in realistic cosmology. Regarding this mechanism we only note that the existence of an (A)dS solution with an arbitrary $\Lambda$ is possible due to the fact that the purely gravitational action (\ref{action}) transforms homogeneously under the global rescaling (\ref{scaling}), $g_{\mu\nu}\to\lambda g_{\mu\nu}$, $S[\lambda g_{\mu\nu}]=\lambda S[g_{\mu\nu}]$, so the crossover mechanism can be based on the breakdown of this property by matter fields, cf. \cite{Shaposhnikovetal}.

The last but not the least is the justification of the choice of (\ref{action}) as an approximation for the effective action coming from some fundamental quantum gravity theory. Quantum effective action with the scaling property of the above type is hard to imagine within semiclassical expansion which contains additional to $M_P^2$ dimensional parameters and has another scaling behavior \cite{quantum0,quantum1}. However, infrared nonlocal expansion in the heat kernel theory of gravitating models and their nonlocal effective action \cite{nneag,TMF} contain nonlocal structures similar to (\ref{action}).

Apart from pragmatic applications in cosmology the model in question can be interesting from pure field-theoretical viewpoint. It is usually considered that modifications of Einstein theory are associated with new degrees of freedom of the gravitational field or broad graviton resonances \cite{DvaliHofmannKhoury}. The above model shows that it is not necessarily the case, because both phases of the theory have the same set of Einstein massless modes. In fact, our model presents the way how the vacuum equation of motion
    \begin{eqnarray}
    R_{\mu\nu}-\Lambda g_{\mu\nu}=0   \label{Einsteinspace}
    \end{eqnarray}
can be encapsulated into the action functional in such a sophisticated way that feeding the theory with matter sources takes place with a nonlocal coupling varying from one of its stable phases to another. Usually this is very hard to implement if we want to maintain stability and unitarity of the theory. Interesting example of such a procedure is a local model of conformal gravity \cite{Maldacena}, in which higher-derivative ghosts are advocated to be eliminated by special boundary conditions. Our model is the example of a nonlocal model explicitly demonstrating absence of ghosts on two vacua of the theory and possible strong and weak coupling regimes associated with these vacua. Moreover, both of these flat and (A)dS vacua have zero value of their on-shell action (\ref{zeroonshell}), which makes the situation very attractive because their contributions to physical amplitudes become equally important without being infinitely suppressed or enhanced, as it happens in other models of Euclidean quantum gravity.\footnote{The degeneracy in the family of (A)dS vacua with different $\Lambda$ might be interpreted as instability with respect to classical or quantum leaps between them. However, classically $\Lambda$ is a constant of motion, whereas its quantum mechanical treatment does not make much sense, because we already work within effective theory for mean fields, which resulted from quantization, and $\Lambda$ is the mean value of the effective cosmological constant.}

In conclusion we mention that serendipity of ghost-free nonlocal gravity models (\ref{action}) might not be exhausted by applications in cosmology. In particular, they might admit generic (not maximally symmetric) Einstein space solutions (\ref{Einsteinspace}) with an arbitrary cosmological constant. Therefore, these models can have implications in black hole physics and be an alternative to the conformal gravity model of \cite{Maldacena}, whereas with a negative $\Lambda$ they become a new testing ground for AdS/CFT correspondence perhaps promising other exciting consequences.

\section*{Acknowledgements}
The author strongly benefitted from thought-provoking criticism of G. Dvali and fruitful discussions with O. Andreev, S. Hofmann, V. Mukhanov, I. Sachs, M. Shaposhnikov and S. Solodukhin.  This work was supported by the Humboldt Foundation at the Physics Department of the Ludwig-Maximilians University in Munich and by the RFBR grant No 11-02-00512.

\section*{NOTE ADDED IN PROOF}
At the proofreading stage of this work the paper \cite{Solodukhin} appeared, advocating that the above results can be generalized to the case of a generic not maximally symmetric Einstein space background, provided that the coefficient of the Riemann term $a$ in Eq.(2) equals $a=2$ (in contrast to the initially expected zero value). This has important implications for a wide class of Schwarzschild-de Sitter black hole solutions with vanishing horizon entropy \cite{Solodukhin}.

\appendix
\renewcommand{\thesection}{\Alph{section}}
\renewcommand{\theequation}{\Alph{section}.\arabic{equation}}

\section{Quadratic part of the action}
Symmetric variation rule for the Green's function (\ref{symvar}) allows one to write down the first order variation of the action (\ref{action}) under the metric variation $\delta g_{\mu\nu}\equiv h_{\mu\nu}$
    \begin{eqnarray}
    &&\delta S=\frac{M^2}2\int d^4x\,g^{1/2}\left\{\,G^{\mu\nu}h_{\mu\nu}
    +\frac\alpha2\,h\,R^{\mu\nu}\frac1{\Box+\hat P} G_{\mu\nu} \right.\nonumber\\
    &&\quad\quad\quad
    +\alpha\,\delta R^{\mu\nu}\frac1{\Box+\hat P}\, G_{\mu\nu}
    +\alpha\,R^{\mu\nu}\frac1{\Box+\hat P}\,\delta G_{\mu\nu}\nonumber\\
    &&\quad\quad\quad\left.
    -\alpha\,R^{\mu\nu} \frac1{\Box+\hat P}\,\big(\delta
    \overrightarrow{\Box}+\delta\hat P\big)
    \frac1{\Box+\hat P}\,G_{\mu\nu}\right\}.  \label{1var}
    \end{eqnarray}
Here and in what follows we use obvious but deserving special mentioning notations
    \begin{eqnarray}
    &&h\equiv g^{\mu\nu}h_{\mu\nu},\quad h^{\mu\nu}\equiv
    g^{\mu\alpha}g^{\nu\beta}h_{\alpha\beta},\quad
    \delta g^{\mu\nu}=-h^{\mu\nu},  \nonumber\\
    &&\delta R^{\mu\nu}= \delta(g^{\mu\alpha}g^{\nu\beta}R_{\alpha\beta})= g^{\mu\alpha}g^{\nu\beta}\delta R_{\alpha\beta}-2h^{\alpha(\mu}R^{\nu)}_\alpha. \nonumber
    \end{eqnarray}
The arrow over $\Box$ indicates the direction in which this operator is acting. For its metric variation $\delta\Box$ it is important to indicate whether it acts to the right or to the left, because in contrast to the symmetric operator $g^{1/2}\Box$ the operator $g^{1/2}\delta\Box$ is not symmetric, and under integration by parts (reversing the direction of the arrow) generates extra terms $\propto\delta g^{1/2}=\frac12g^{1/2}h$.

The expression (\ref{1var}) allows one to derive explicitly the variational derivative of the action on the (A)dS background and confirm Eqs.(\ref{500})-(\ref{501}) derived above by the conformal variation method. Using (\ref{GreenfunconR}) and (\ref{GreenfunconG})
and reversing the action of the symmetric $(\Box+\hat P)^{-1}$ to the left in the fourth and fifth terms of (\ref{1var}) we have
    \begin{eqnarray}
    &&\delta S\,\Big|_{\,\rm (A)dS}=\frac{M^2}2\int d^4x\,g^{1/2}\left\{\,-\Lambda h
    -\frac{2\alpha\Lambda}{A+4B}\,h\right.\nonumber\\
    &&\qquad\quad\qquad
    -\frac\alpha{A+4B}\,g_{\mu\nu}\delta R^{\mu\nu}
    +\frac\alpha{A+4B}\,g^{\mu\nu}\,\delta G_{\mu\nu}\nonumber\\
    &&\qquad\quad\qquad\left.
    +\frac\alpha{(A+4B)^2}\,g^{\mu\nu}\,\big(\delta
    \overrightarrow{\Box}
    +\delta\hat P\big)\,g_{\mu\nu}\right\}.  \label{1varAdS}
    \end{eqnarray}
To calculate the first four terms here we use the Ricci tensor variation
    \begin{eqnarray}
    &&\delta R_{\mu\nu}=
    \nabla^\alpha\nabla_{(\mu}h_{\nu)\alpha}
    -\frac12\,\Box h_{\mu\nu}
    -\frac12\,\nabla_\mu\nabla_\nu h,    \label{Riccivariation}
    \end{eqnarray}
while the fifth term can be obtained as follows. Consider the first order metric variation of the identities $\Box g_{\mu\nu}=0$ and (\ref{Pg}). This gives
    \begin{eqnarray}
    &&(\delta\overrightarrow{\Box})\,
    g_{\mu\nu}=-\Box h_{\mu\nu},   \label{deltaBoxong}\\
    &&(\delta\hat P)\,
    g_{\mu\nu}=-\hat P\, h_{\mu\nu}
    +A\,\delta R_{\mu\nu}
    +B\,R\,h_{\mu\nu}\nonumber\\
    &&\qquad\qquad\quad
    +B\,g_{\mu\nu}\,\delta R,     \label{deltaPong}
    \end{eqnarray}
so that on the (A)dS background
    \begin{eqnarray}
    &&\big(\delta\overrightarrow{\Box}+\delta\hat P\big) g_{\mu\nu}=-\big(\overrightarrow{\Box}+\hat P\big) h_{\mu\nu}+A\,\delta R_{\mu\nu}
    \nonumber\\
    &&\qquad\qquad\qquad\qquad
    +4B\Lambda\,h_{\mu\nu}
    +B\,g_{\mu\nu}\,\delta R.       \label{deltaBoxdeltaPong}
    \end{eqnarray}
Therefore the variation (\ref{1varAdS}) takes the form
    \begin{equation}
    \delta S\,\Big|_{\,\rm (A)dS}=-\frac{M^2\Lambda}2\left(1+\frac\alpha{A+4B}\right)
    \int d^4x\,g^{1/2} g^{\mu\nu}\delta g_{\mu\nu}
    \end{equation}
and confirms Eqs.(\ref{500})-(\ref{501}) which lead to the criterion (\ref{relation}) of the (A)dS background with an arbitrary $\Lambda$.

Now let us go over to the quadratic part of the action on this background. For this we make the metric variation of (\ref{1var}) by applying the same rule (\ref{symvar}) and integration by parts
    \begin{eqnarray}
    &&S_{(2)}=\frac12\,\delta^2S\,\Big|_{\,\rm (A)dS}.
    \end{eqnarray}
The result immediately simplifies if we use the following two corollaries
    \begin{eqnarray}
    &&g^{\mu\nu}\frac1{\Box+\hat P}\,\Phi_{\mu\nu}
    =\frac1{\Box-\alpha\Lambda}\,(g^{\mu\nu}\Phi_{\mu\nu}),\\
    &&\frac1{\Box+\hat P}\,(g^{\mu\nu}\Phi)
    =g^{\mu\nu}\frac1{\Box-\alpha\Lambda}\,\Phi
    \end{eqnarray}
of the equation
    \begin{eqnarray}
    &&g^{\alpha\beta}\big(\Box+\hat P\big)_{\alpha\beta}^{\;\;\;\mu\nu}
    =\big(\Box-\alpha\Lambda\big)\,g^{\mu\nu},
    \end{eqnarray}
which is of course based on the relations (\ref{Pg1}) and (\ref{relation}). The above equations hold for generic tensor $\Phi^{\mu\nu}$ and scalar $\Phi$ fields and allow us to pull the metric tensor through the tensor propagator converting it into the scalar one.

As a result the quadratic part of the action splits into the sum
    \begin{eqnarray}
    &&S_{(2)}=\overline S_{(2)}+\widetilde S_{(2)}
    \end{eqnarray}
of the purely local term
    \begin{eqnarray}
    &&\overline S_{(2)}=\frac{M^2}4\int d^4x\,g^{1/2}\left\{\,\frac12\,h\,\delta R+\frac12\,\Lambda\,h^2\right.\nonumber\\
    &&\qquad\quad\quad
    +\delta G^{\mu\nu}h_{\mu\nu}
    -\frac12\,h_{\mu\nu}^2R
    +\frac12\,h\,g_{\mu\nu}\delta R^{\mu\nu}
    \nonumber\\
    &&\qquad\quad\quad
    +\big(\delta^2 R^{\mu\nu}\big)\,g_{\mu\nu}-g^{\mu\nu}\big(\delta^2 G_{\mu\nu}\big)
    \nonumber\\
    &&\qquad\quad\quad\left.
    +\frac1\alpha\,g^{\mu\nu}
    \big(\delta^2\overrightarrow{\Box}
    +\delta^2\hat P\big)\, g_{\mu\nu}\right\} \label{barS}
    \end{eqnarray}
and the term which together with some local pieces contains all nonlocalities of $S_{(2)}$
    \begin{eqnarray}
    &&\widetilde S_{(2)}=\frac{M^2}4\int d^4x\,g^{1/2}\left\{\,
    +\frac{\alpha\Lambda}2\,h\,\frac1{\Box-\alpha\Lambda}\,g^{\mu\nu}\delta G_{\mu\nu}\right.
    \nonumber\\
    &&\qquad\quad\qquad
    -\frac\Lambda2\,h\,\frac1{\Box-\alpha\Lambda}\,
    g^{\mu\nu}\big(\delta\overrightarrow{\Box}+\delta\hat P\big)\, g_{\mu\nu}
    \nonumber\\
    &&\qquad\quad\qquad
    -\frac2\alpha g^{\mu\nu}\big(\delta\overrightarrow{\Box}+\delta\hat P\big) \frac1{\Box+\hat P}\,\big(\delta\overrightarrow{\Box}+\delta\hat P\big) g_{\mu\nu}\nonumber\\
    &&\qquad\quad\qquad
    +2\,\alpha\,\delta R^{\mu\nu}\frac1{\Box+\hat P}\,\delta G_{\mu\nu}\nonumber\\
    &&\qquad\quad\qquad
    -2\,\delta R^{\mu\nu}\frac1{\Box+\hat P}\,\big(\delta\overrightarrow{\Box}+\delta\hat P\big) g_{\mu\nu}\nonumber\\
    &&\qquad\quad\qquad\left.
    +2\,g^{\mu\nu}\big(\delta\overrightarrow{\Box}
    +\delta\hat P\big)
    \frac1{\Box+\hat P}\,\delta G_{\mu\nu}\right\}. \label{tildeS}
    \end{eqnarray}

\subsection{Disentangling nonlocal structures}
Here we begin with disentangling nonlocal terms from (\ref{tildeS}). From now on we will consider metric perturbations in the DeWitt gauge (\ref{DWgauge}) in which the Ricci tensor variation (\ref{Riccivariation}) (on the (A)dS-background) simplifies to
    \begin{equation}
    \delta R_{\mu\nu}\Big|_{\,\rm (A)dS}=-\frac12\,\Box h_{\mu\nu}+\frac43\,\Lambda\left(h_{\mu\nu}
    -\frac14\,g_{\mu\nu}h\right).           \label{deltaRicci}
    \end{equation}
Therefore, the expression (\ref{deltaBoxdeltaPong}) takes the form
    \begin{eqnarray}
    &&g^{\mu\nu}\big(\delta\overrightarrow{\Box}+\delta\hat P\big) g_{\mu\nu}=-\left(1-\frac\alpha2\right)\,
    (\Box-\alpha\Lambda)\,h\nonumber\\
    &&\qquad\qquad\qquad\qquad\qquad
    +\frac{\alpha^2}2\,\Lambda\,h.      \label{gdeltaBoxdeltaPong}
    \end{eqnarray}

Using this expression and the fact that $g^{\mu\nu}\delta G_{\mu\nu}=\Box h/2$ one finds that the sum of the first two nonlocal terms in (\ref{tildeS}) is in fact local (here we consider all the contributions to (\ref{tildeS}) modulo the overall factor $M^2/4$)
    \begin{eqnarray}
    &&\!\!\!\!\!\!\!\!\!\!\!\!\!\!\!\!\int d^4x\,g^{1/2}\,\left\{
    -\frac\Lambda2\,h\,\frac1{\Box-\alpha\Lambda}\,
    g^{\mu\nu}\big(\delta\overrightarrow{\Box}+\delta\hat P\big)\, g_{\mu\nu}\right.
    \nonumber\\
    &&\!\!\!\!\!\!\!\!\left.+\frac{\alpha\Lambda}2\,h\,
    \frac1{\Box-\alpha\Lambda}\,
    g^{\mu\nu}\delta G_{\mu\nu}\right\}=
    \frac\Lambda2\int d^4x\,g^{1/2}\,h^2.   \label{A99}
    \end{eqnarray}

Using the relation (\ref{deltaBoxdeltaPong}) and its transpose\footnote{Functional transposition of $\delta{\Box}$ should take into account the variation of the integration measure $g^{1/2}$ because, as mentioned above, the operator $g^{1/2}(\delta\Box)$ is not symmetric in contrast to symmetric $g^{1/2}\Box$.} one can show that the last three terms in (\ref{tildeS}) reduce to
    \begin{eqnarray}
    &&\int d^4x\,g^{1/2}\left\{\,
    +2\,\alpha\,\delta R^{\mu\nu}\frac1{\Box+\hat P}\,\delta G_{\mu\nu}\right.\nonumber\\
    &&\qquad\qquad\qquad
    -2\,\delta R^{\mu\nu}\frac1{\Box+\hat P}\,\big(\delta\overrightarrow{\Box}+\delta\hat P\big) g_{\mu\nu}\nonumber\\
    &&\qquad\qquad\qquad
    \left.+2\,g^{\mu\nu}\big(\delta\overrightarrow{\Box}+\delta\hat P\big) \frac1{\Box+\hat P}\,\delta G_{\mu\nu}\right\}\nonumber\\
    &&=\int d^4x\,g^{1/2}\left\{\,\vphantom{\frac11}4h^{\mu\nu}\delta R_{\mu\nu}-8\Lambda h_{\mu\nu}^2+\Lambda\left(1-\frac\alpha2\right) h^2\right.\nonumber\\
    &&%\qquad\qquad\qquad\quad
    \;\;+2\,\alpha\,\delta R^{\mu\nu}\frac1{\Box+\hat P}\,\delta R_{\mu\nu}%\nonumber\\
    %&&\qquad\qquad\qquad\quad
    \left.
    -\frac{\alpha^2\Lambda^2}2\,h\,
    \frac1{\Box-\alpha\Lambda}\,h\,\right\}.     \label{A100}
    \end{eqnarray}
Finally, the third term in (\ref{tildeS}) can be transformed as
    \begin{eqnarray}
    &&-\frac2\alpha\int d^4x\,g^{1/2}\,
     g^{\mu\nu}\big(\delta\overrightarrow{\Box}+\delta\hat P\big)
     \frac1{\Box+\hat P}\,\big(\delta\overrightarrow{\Box}+\delta\hat P\big) g_{\mu\nu}\nonumber\\
    &&\qquad
    =\frac1{2\alpha}\int d^4x\,g^{1/2}\,\left\{\vphantom{\frac11}\Big[\,\alpha-2
    -2B(A+2B)\,\Big]\,h\,\Box h\right.\nonumber\\
    &&\qquad+4\,h^{\mu\nu}\Box h_{\mu\nu}
    +4\Lambda h_{\mu\nu}^2\Big[\;2\alpha-C\,\Big]\nonumber\\
    &&\qquad+\Lambda h^2\left[\,C
    -B(A+2B)\left(8+2\alpha\right)-\frac\alpha2\,\right]\nonumber\\
    &&\qquad+\Lambda^2\Big[\,\alpha^3-
    B(A+2B)\left(8\alpha+2\alpha^2+8\right)\Big]\,
    h\frac1{\Box-\alpha\Lambda}h\nonumber\\
    &&\qquad+32\Lambda^2 B(A+2B)\,h^{\mu\nu}
    \frac1{\Box+\hat P}\,h_{\mu\nu}\nonumber\\
    &&\qquad\left.-4A^2\,
    \delta R^{\mu\nu}
    \frac1{\Box+\hat P}\,\delta R_{\mu\nu}\right\}.   \label{A101}
    \end{eqnarray}

To calculate $\delta R^{\mu\nu}\times\delta R_{\mu\nu}$-terms in the above two expressions we use another form of the Ricci tensor variation which holds in view of (\ref{deltaRicci}) and (\ref{PonAdS})
    \begin{eqnarray}
    &&\delta R_{\mu\nu}\big|_{\,\rm (A)dS}=-\frac12\,\big(\Box+\hat P\big) h_{\mu\nu}-\frac\alpha8\Lambda\,g_{\mu\nu}h\nonumber\\
    &&\qquad\quad\quad
    -\Lambda\left(\,\frac{C}2-\frac43\,\right)\,
    \Big(h_{\mu\nu}-\frac14\,g_{\mu\nu}h\Big).
    \end{eqnarray}
This form contains as a whole the operator $\Box+\hat P$ which cancels $(\Box+\hat P)^{-1}$ and renders a part of terms explicitly local. The result reads
    \begin{eqnarray}
    &&\!\!\!\!\!\!\!\!\int d^4x\,g^{1/2}\,
    \delta R^{\mu\nu}
    \frac1{\Box+\hat P}\,\delta R_{\mu\nu}\nonumber\\
    &&\quad=\int d^4x\,g^{1/2}\,\left\{\,\frac14 h^{\mu\nu}\Box h_{\mu\nu}
    +\frac\Lambda4 h_{\mu\nu}^2\Big(\,C-\frac43\,\Big)\right.\nonumber\\
    &&\quad+\frac\Lambda{16}\, \Big(\,\alpha-C+\frac{16}3\,\Big)\,h^2\nonumber\\
    &&\quad+\frac{\Lambda^2}{16}\,
    \Big(\,\alpha-C+\frac83\,\Big)\Big(\,\alpha+C+\frac43\,\Big)\,
    h\,\frac1{\Box-\alpha\Lambda}\,h\nonumber\\
    &&\quad\left.+\frac{\Lambda^2}4\,
    \Big(\,C+\frac43\,\Big)\,
    \Big(\,C-\frac83\,\Big)\,h^{\mu\nu}
    \frac1{\Box+\hat P}\,h_{\mu\nu}\right\}.   \label{A103}
    \end{eqnarray}
Using this relation and collecting in (\ref{A100}) and (\ref{A101}) the coefficients of two nonlocal structures we get the nonlocal part of (\ref{tildeS}). In particular,
the overall coefficient of the $h^{\mu\nu}(\Box+\hat P)^{-1}h_{\mu\nu}$ nonlocality turns out to be
    \begin{eqnarray}
    &&\frac{\Lambda^2}{2\alpha}\,\left(C+\frac43\right)
    \left(C-\frac83\right)\,(\alpha^2-A^2)
    +\frac{16\Lambda^2}{\alpha}\,B(A+2B)\nonumber\\
    &&\qquad\qquad\qquad=
    \frac{4B(A+2B)}\alpha\,\Lambda^2\left(C-\frac23\right)^2,
    \end{eqnarray}
because $\alpha^2-A^2=8B(A+2B)$. Similarly the coefficient of $h(\Box-\alpha\Lambda)^{-1}h$ equals
    \begin{eqnarray}
    &&\frac{\Lambda^2}{8\alpha}\,\left(\alpha+C+\frac43\right)
    \left(\alpha-C+\frac83\right)\,(\alpha^2-A^2)\nonumber\\
    &&
    -\Lambda^2\,B(A+2B)\left(4+\alpha+\frac4\alpha\right)\nonumber\\
    &&\qquad\quad\quad=
    -\frac{B(A+2B)}\alpha\,\Lambda^2\left(C-\frac23\right)^2.
    \end{eqnarray}
This finally gives the nonlocal terms of Eq.(\ref{s_2}) and explains the origin of the effective Planck mass (\ref{effectivemass}).

\subsection{Local terms}
Local terms of the quadratic part of the action are contained both in (\ref{barS}) and (\ref{tildeS}). The most economical way to simplify the second variations of Ricci and Einstein tensors in (\ref{barS}) is to use the identity $(\delta^2 R^{\mu\nu})\,g_{\mu\nu}-g^{\mu\nu}(\delta^2 G_{\mu\nu})=\delta(\delta R^{\mu\nu} g_{\mu\nu}-g^{\mu\nu}\delta G_{\mu\nu})-h_{\mu\nu}\delta R^{\mu\nu}-h^{\mu\nu}\delta G_{\mu\nu}$ and then repeat this trick for first order variations. In this way the second order variations reduce to the calculation of
    \begin{eqnarray}
    &&\int d^4x\,g^{1/2}\,\delta^2R=\int d^4x\,g^{1/2}\,\left\{\,2\Lambda h_{\mu\nu}^2-\frac12\,\Lambda h^2\right.\nonumber\\
    &&\qquad\qquad\qquad\qquad\left.
    -h^{\mu\nu}\delta R_{\mu\nu}-\frac12\,h\,\delta R\right\},
    \end{eqnarray}
and the first three lines of (\ref{barS}) simplify to
    \begin{eqnarray}
    &&\int d^4x\,g^{1/2}\left\{\,\frac12\,h\,\delta R+\frac12\,\Lambda\,h^2+\delta G^{\mu\nu}h_{\mu\nu}\right.\nonumber\\
    &&\qquad\qquad\quad
    -\frac12\,h_{\mu\nu}^2R
    +\frac12\,h\,g_{\mu\nu}\delta R^{\mu\nu}
    \nonumber\\
    &&\qquad\qquad\quad\left.
    +\big(\delta^2 R^{\mu\nu}\big)\,g_{\mu\nu}-g^{\mu\nu}\big(\delta^2 G_{\mu\nu}\big)\vphantom{\frac11}\right\}\nonumber\\
    %&&=\int d^4x\,g^{1/2}\left\{-\frac14\,h\Box h+4\Lambda %h_{\mu\nu}^2-\Lambda h^2+\frac12\,h\,\delta R\right.\nonumber\\
    %&&\qquad\qquad\qquad\quad
    %\left.+\frac12\,h g^{\mu\nu}\delta R_{\mu\nu}-3h^{\mu\nu}\delta %R_{\mu\nu}+2\delta^2 R\right\}\nonumber\\
    &&=\int d^4x\,g^{1/2}\left\{\frac52\,h^{\mu\nu}\Box h_{\mu\nu}
    -\frac14\,h\,\Box h\right.\nonumber\\
    &&\qquad\qquad\qquad\qquad\quad
    \left.+\frac43\,\Lambda
    h_{\mu\nu}^2+\frac16\,\Lambda h^2\right\}.    \label{B1}
    \end{eqnarray}

For the calculation of the last term of (\ref{barS}) consider the second order variation of the identities $\Box g_{\mu\nu}=0$ and (\ref{Pg})
    \begin{eqnarray}
    &&(\delta^2\overrightarrow{\Box})\,
    g_{\mu\nu}=
    -2(\delta\overrightarrow{\Box})\,
    h_{\mu\nu},                          \label{delta2boxong}\\
    &&(\delta^2\hat P)\,
    g_{\mu\nu}=-2\delta\hat P\, h_{\mu\nu}
    +A\,\delta^2 R_{\mu\nu}
    +2\,B\,\delta R\,h_{\mu\nu}\nonumber\\
    &&\qquad\qquad\quad
    +B\,g_{\mu\nu}\,\delta^2 R,              \label{delta2Pong}
    \end{eqnarray}
and again use the first order variations  (\ref{deltaBoxong})-(\ref{deltaPong}). Then bearing in mind that $g^{\mu\nu}(\delta\Box) h_{\mu\nu}=\delta(g^{\mu\nu}\Box)h_{\mu\nu}+h^{\mu\nu}\Box h_{\mu\nu}=(\delta\Box)h-\Box(h_{\mu\nu}^2)+h^{\mu\nu}\Box h_{\mu\nu}$ we reduce the last term of $\overline S_{(2)}$ to
    \begin{eqnarray}
    &&\frac1{\alpha}\int d^4x\,g^{1/2}\,g^{\mu\nu}\big(\delta^2\overrightarrow{\Box}
    +\delta^2\hat P\big)\, g_{\mu\nu}
    \nonumber\\
    &&\qquad=\frac1{\alpha}\int d^4x\,g^{1/2}\left\{\vphantom{\frac11}\,h\,\Box
    h-2h^{\mu\nu}\Box
    h_{\mu\nu}-\alpha\,\delta^2R\right.\nonumber\\
    &&\quad\qquad\qquad\qquad\left.+2\Lambda\left(\,C-\alpha\,\right)\,
    \Big(h_{\mu\nu}^2-\frac14\,g_{\mu\nu}h^2\Big)
    \right\}\nonumber\\
    &&\qquad=\frac1{\alpha}\int d^4x\,g^{1/2}\left\{\left(1-\frac\alpha4\right) h\Box h-\left(2+\frac\alpha2\right) h^{\mu\nu}\Box h_{\mu\nu}\vphantom{\frac11}\right.\nonumber\\
    &&\quad\qquad\quad\left.
    +\Lambda h_{\mu\nu}^2\Big(2C-\frac{8\alpha}3\Big)
    +\frac\Lambda2\,h^2
    \Big(\frac\alpha3-C\Big)\right\}.     \label{B2}
    \end{eqnarray}

Then collecting the local terms from (\ref{A99})-(\ref{A101}) with the $\delta R^{\mu\nu}\times\delta R_{\mu\nu}$-terms given by (\ref{A103}) together with the local terms (\ref{B1}) and (\ref{B2}) we finally get the local part of (\ref{s_2}).

\end{document}